# On the nature of the (de)coupling of the magnetostructural transition in $Er_5Si_4$

Rui M. Costa[1], João H. Belo[1], Marcelo B. Barbosa[1], Pedro A. Algarabel[2,3], César Magén[2,3,4], Luis Morellón[3,5], Manuel R. Ibarra[2,5], João N. Gonçalves[6], Nuno M. Fortunato[6], João S. Amaral[6], João P. Araújo[1], and André M. Pereira[*,1]

[1] IFIMUP and IN-Institute of Nanoscience and Nanotechnology, Porto University, Rua do Campo Alegre, 687, 4169-007 Porto, Portugal

[2] Instituto de Ciencia de Materiales de Aragón, Universidad de Zaragoza and Consejo Superior de Investigaciones Científicas, 50009 Zaragoza, Spain

[3] Depatamento de Física de la Materia Condensada, Universidad de Zaragoza, 50009 Zaragoza, Spain

[4] Instituto de Nanociencia de Aragón-ARAID, Universidad de Zaragoza, 50009 Zaragoza, Spain

[5] Instituto de Nanociencia de Aragón, Universidad de Zaragoza, 50009 Zaragoza, Spain

[6] Departamento de Física e CICECO, Universidade de Aveiro, 3810-193 Aveiro, Portugal

[*] Corresponding author: ampereira@fc.up.pt

**Abstract**

In this report, a successful thermodynamical model was employed to understand the structural transition in $Er_5Si_4$, able to explain the decoupling of the magnetic and structural transition. This was achieved by DFT calculations, which were used to determine the energy differences at 0 K, using a LSDA+U approximation. It was found that the M structure is the stable phase at low temperatures, as verified experimentally with a value of $\Delta F_0$= -0.262 eV. Finally, a variation of Seebeck coefficient (~6 µV) was determined at the structural transition, which allows to conclude that the electronic entropy variation is negligible in the transition.

# 1. Introduction

The magnetocaloric effect (MCE) can be defined as the entropy change in an isothermal process or as the temperature change in an adiabatic process due to an applied magnetic field and is at the basis of the magnetic refrigeration technology [1–3].

This subject gained more interest with the discovery of the giant MCE (GMCE) near room temperature, making magnetic refrigeration possible in conventional household appliances. These high values arise from a first-order phase transition (FOPT) that in many cases comprises a magnetostructural transition, i.e., a coupled magnetic and structural transition, such as $R_5(Si,Ge)_4$, MnAs, MnFe(As,P), FeRh, among others [2].

From the families exhibiting GMCE, the $R_5(Si,Ge)_4$ presents one of the highest MCE and a wide temperature range where the effect is present (20 – 335 K) [2].

The compounds from this family have 36 atoms: 20 R (where R = rare earth) and 16 Si per unit cell; and can be viewed as made of rigid slabs stacked on top of each other where, depending on the bonds between these slabs, the system can present one of the following crystallographic structures: orthorhombic (O(I)) (Pnma space group) with all interslab bonds formed, monoclinic (M) ($P112_1/a$) with half of the interslab bonds formed, and orthorhombic (O(II)) (Pnma) with no interslab bonds [3]. Also, nearly every member that stabilizes in the M or O(II) structure and in the paramagnetic state (PM), goes through a first-order martensitic-like structural transformation to O(I) and becomes ferromagnetic (FM) while cooling [4], such as $Gd_5Si_2Ge_2$ [5], $Tb_5Si_2Ge_2$ [6], $Dy_5Si_3Ge$ [7], etc.

The only system that does not present this behaviour is the $Er_5Si_4$. In this system, the structural transition, $T_S$ ~ (200 - 230) K [4], follows from O(I) to M on cooling which is in fact the stable structure expected since a reduction of symmetry is common when lowering the temperature for systems in equilibrium. Magnetically, it presents an ordering temperature of $T_C^M \approx 30$ K from PM to a canted ferromagnetic ordering with an easy-axis along the *b*-axis [8].

One of the first theoretical works covering this system used the density of states (DOS) and crystal orbital Hamilton/overlap population (COHP/COOP) calculated from first principles [9]. The conclusions were that, in this structural transition, the shear movement of the slabs, comes as a response to the reduction of itinerant electrons since it enhances the Si–Si bonds but weakens the remaining bonds. Moreover, pressure revealed to tune the magnetostructural transition allowing a fully coupled magnetic and structural transition to occur simultaneously [10]. Although several studies were devoted to this material, an assessment of the thermodynamic nature of this structural transition is still in pursuit.

Paudyal et al. [11] developed a thermomagnetic model, where the electronic calculations are coupled to a mean field model. This model was able to successfully explain, for the first time in the R$_5$(Si,Ge)$_4$ family, the behavior of Gd$_5$Si$_2$Ge$_2$ [11]. They were able to determine the transition temperature due to the crossing of the M and O(I) free energies and the finite temperature properties, e.g., magnetization, magnetic free energy, and magentic entropy change, which find good agreement with the experimental results. More recently, this same model was employed to investigate Tb$_5$Si$_2$Ge$_2$, correctly predicting the structural transition and its slight decoupling with the magnetic transition [12].

In summary, the model assumes that there is a competition between two phases and that the free energy of each system can be divided in three parts [1]:

$$F(T,B) = F_M(T,B) + F_L(T) + F_0 \tag{1}$$

where $T$ and $B$ are the temperature and applied induction field, respectively, $F_M$ is the magnetic contribution, $F_L$ is the lattice term and $F_0$ is the free energy at 0 K. It is assumed that the magnetic component is described by a Mean Field Approximation (MFA) where each structure presents a different $T_C$ [13]:

$$F_M(T,) = -N_M k_B T \ln\left(\frac{\sinh\left(\frac{2J+1}{2J}\frac{3J}{J+1}\left(\frac{B}{\lambda M_S}+\sigma(T,B)\right)\frac{T_C}{T}\right)}{\sinh\left(\frac{1}{2J}\frac{3J}{J+1}\left(\frac{B}{\lambda M_S}+\sigma(T,B)\right)\frac{T_C}{T}\right)}\right) \tag{2}$$

where $N_M$ is the number of magnetic ions, $k_B$ is the Boltzmann constant, $J$ is the total angular momentum of the magnetic ions, $\lambda$ is the constant that parametrizes the intensity of the molecular field, $M_S$ is the saturation magnetization, $\sigma$ is the normalized magnetization, and $T_C$ is the Curie temperature. The lattice term is obtained from the Debye model resulting in:

$$F_L(T) = \frac{9Nk_B\Theta_D}{8} + 3Nk_BTln(1 - e^{-\Theta_D/T}) - 3Nk_BT\left(\frac{T}{\Theta_D}\right)^3 \int_0^{-\Theta_D/T} \frac{x^3}{e^x - 1}dx \qquad (3)$$

where $N$ is the number of ions and $\Theta_D$ is the Debye temperature. Finally, due to the electronic structure complexity, the free energy at 0 K, $F_0$, is obtained from density functional theory (DFT) based on first-principles.

## 2. Methodology

The magnetization temperature dependence in the range 5-300 K and isothermal high magnetic field magnetization measurements for temperatures between 5 and 215 K considered in this work were extracted from Refs. [14] and [15], respectively.

For the computation of the free energy at 0 K, $F_0$, DFT calculations were performed using the Linearized / Augmented Planewave plus local orbitals (LAPW+lo) method implemented in the WIEN2k package [16]. The GGA+U approach has been used to study the electronic and magnetic properties of $Er_5Si_4$. The functional adopted for the exchange correlation interaction (GGA) is the one given by Perdew-Burke-Ernzerhof [17]. To describe the 4$f$ electrons of the Er atoms an effective Coulomb interaction of U = 5.8 eV was chosen [18, 19]. The main idea behind the GGA+U is that, since GGA misrepresents the strongly correlated states, a correction described by the Hubbard model is added to those states in order to improve it. To ensure a good accuracy for the free energy model, a $10^{-4}$ Ry per unit cell was chosen as the convergence criterion. Furthermore, 126 and 168 k-points in the irreducible part of the first Brillouin zone were selected for the O(I) and M structures, respectively, and a cutoff parameter of $R_{km}K_{max} = 7$, where $R_{km}$ is the smallest muffin tin radius and $K_{max}$ is the maximal value of the wavevector of the plane wave expansion. Additionally, these computations were performed with spin polarization. Further complementary DFT calculations were performed using the Viena Ab initio simulation Package (VASP) [20], including spin-polarized calculations for GGA+U values between 0 and 12 eV and simulations with spin-orbit coupling at U=5.8 eV. The energy cut-off for the plane waves was set at 388 eV and a smaller set of 36 k-points in the irreducible Brillouin zone was found to be sufficient to converge the relaxed lattice parameters to under 0.5%. Both lattice parameters and magnetization values were in agreement with the more precise WIEN2K simulations, used to determine the free energy difference between structures.

## 3. Electronic band structure

The energy differences between the M and O(I) was computed with and without the Hubbard parameter resulting in 0.486 and -0.262 eV respectively, after minimizing each structure. These results show that only by adding the Hubbard parameter, U, the M structure presents lower energy than the O(I) implying that it is the stable phase at T = 0 K, which is consistent with the experimental results. This confirms the strong electronic correlations present in this system and reflects the need to include the parameter in the simulations as seen in Ref. [21]. It was also observed that after optimization and minimization of the atomic positions (see Table 1), the structure obtained is very similar to the one measured experimentally [15] suggesting that the value of 5.8 eV for the Hubbard parameter is reliable.

Although the low and high temperature structures in $Er_5Si_4$ are switched relative to $Gd_5Si_2Ge_2$ and $Tb_5Si_2Ge_2$, the magnitude of its energy difference is of the same order as in the latter compounds, 0.36 and 0.11 eV/unit cell respectively [11, 12].

Moreover, the estimated total magnetic moment of the unit cell is 62.569 µB/cell and 65.395 µB/cell for the M and O(I) respectively, corroborating that in the O(I) phase, Si-Si bonds enable long-range Ruderman-Kittel-Kasuya-Yosida (RKKY) ferromagnetic interactions between Er-4*f* moments across slabs as previously observed in $Gd_5Si_2Ge_2$ [22]. Thus, a higher $T_C$ is expected for the O(I) phase in comparison with the M phase.

Even though the results obtained for each structure are comparable, the obtained saturation magnetization is of 96.25 emu g$^{-1}$, which is approximately half of the experimental value, 200 emu g$^{-1}$ [4, 23, 24]. The magnetic moments for the Er atoms, despite being dependent on the sphere radius, are approximately 3µB instead of the 8µB obtained experimentally [8, 15]. In order to better understand this, calculations for several U were performed and the magnetization seemed to reach an asymptotic value of ~68µB. Also, simulations with spin-orbit coupling to account for anisotropies were unsuccessful in explaining this discrepancy, therefore, a more detailed analysis with non-collinear magnetism is needed.

**Table 1** Lattice parameters obtained from the optimization of lattice parameters and minimization of atomic positions. Experimental results for O(I) (at 293 K) and M (at 2 K) were obtained from Refs. [4] and [15] respectively.

|  | WIEN2k |  | Exp. |  |
| --- | --- | --- | --- | --- |
|  | O(I) | M | O(I) | M |
| *a* (Å) | 7.320041 | 7.421816 | 7.28386 | 7.34833 |
| b (Å) | 14.433021 | 14.420224 | 14.3631 | 14.3491 |
| *c* (Å) | 7.632085 | 7.579240 | 7.59436 | 7.54153 |
| $\gamma$ (°) | 90 | 93.2204 | 90 | 93.2204 |
| V (Å$^3$) | 806.3231 | 809.8816 | 794.495 | 793.93 |

The density of states (DOS) was also determined to study the differences in the electronic and magnetic properties (see Fig. 1). The Fermi energy, $E_F$, lies in the conduction band indicating that it is metallic behavior as expected. The valence band goes from -4.5 up to -0.5 eV, the O(I) has a band gap of ~ 0.1 eV for spin up at -0.8 eV and a band gap of ~ 0.2 eV for spin down at -0.55 eV. In contrast, the M structure presents smaller band gaps, 0.03 eV for spin up and 0.05 eV for spin down. These results are consistent with TB LMTO-ASA calculations obtained in Ref. [9]. The valence band is mainly a mixture of Er-3$d$ and Si-2$p$ states while the conduction band is dominated by the Er-3$d$ states in both structures. The Er-4$f$ bands (see Figs. 1a) are extremely localized as expected from the Hubbard term in the exchange functional which is the main difference from the result obtained in Ref. [9]. The appearance of additional peaks below -5 eV in M are associated with the Er1A atoms in the $f$ bands (not shown).

Every Er atom in the M phase has an additional peak just below $E_F$, at -0.4 eV (see Fig. 1a), and this is a direct consequence of the breaking of half the Si-Si dimers. Moreover, the first peak is steeper for Er1A and Er2B (see Ref. [4] for notation) because these atoms are adjacent to the non-bonded Si-Si dimers. On the other hand, O(I) only has the peak at -0.1 eV. The absence of the first peak in the O(I) implies that in this energy range the states can only be spin up polarized, additionally, the spin up DOS of Er-5$d$ of the O(I) at $E_F$ are bigger than for M while the spin down DOS of Er-5$d$ are the same in both structures, which overall suggests that the magnetization for the O(I) structure should be higher.

Another effect of the dimer breaking are the peaks in Si1B-$p$ for M (see Fig. 1b). These are mainly due to Si1B-$p_y$ bands (not shown) and come from the stronger interaction between the Si1B-$p$ and Er3-$d$ due to a decrease in distance, $\delta_{\text{Er3-Si1B}} = 2.96449$ Å. This is

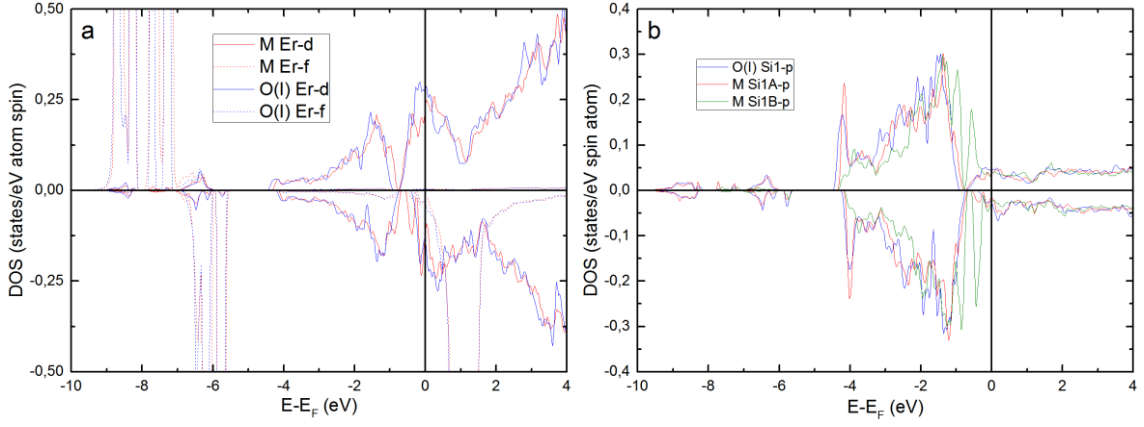

**Figure 1** (a) Average spin up and down DOS of all Er atoms of bands *d* and *f*. (b) Spin up and down DOS of bands *p* of Si1.

not manifested for Si1A because its distance to Er3, $\delta_{Er3\text{-}Si1A} = 3.05646$ Å, is similar as in the case of O(I), $\delta_{Er3\text{-}Si1} = 3.04203$ Å.

Through the use of *ab initio* calculations it was estimated the Seebeck coefficient above $T_C$ and near $T_S$ from [25]:

$$S_D = -\frac{\pi^2 k_B^2 T}{3e}\left(\frac{d\ln\sigma(E,T)}{dE}\right)_{E=E_F} \quad (4)$$

where $k_B$ is the Boltzmann constant, $T$ is the temperature, $e$ is the electronic charge and $\sigma$ is the electrical conductivity. It was assumed that the conductivity can be expressed by $\sigma(E_F) \sim 1/N_d(E_F)$, where $N_d$ is the DOS of the *d* bands, as in the case of $s - d$ scattering. The difference obtained in Seebeck coefficients of both structures at 200 K was $|S_D^{O(I)} - S_D^M| \approx 6$ µV/K, which is in agreement with the experimental result that is in between ~ 6 and 7 µV/K [14]. Additionally, the electronic entropy change can be estimated from [26]:

$$\Delta S_E = \Delta S_D N e \quad (5)$$

where $\Delta S_D$ is the previously calculated Seebeck coefficients difference, $N$ is the carrier concentration and $e$ is the electronic charge. Considering that the Er ions are in the state $Er^{3+}$, the percentage of electrons in the conduction band is ~ 4.4 %. This results in an electronic entropy change, $\Delta S_E$, of ~ 0.079 meV/cell, which is negligible when compared with the $\Delta S_L$ (~ 12 meV/cell, see below).

## 4. Magnetic results

In Fig. 2a the magnetization and inverse susceptibility in the 5-300 K temperature are shown. $Er_5Si_4$ exhibits, as expected, a step-like behavior in the inverse susceptibility from ~200 to ~230 K. This deviation from the Curie-Weiss law signals the presence of a structural transition [4]. On cooling, the magnetic ordering sets in at 30 K which corresponds to the Curie temperature of the M phase, $T_C^M$.

In Fig. 2b the isothermal magnetizations up to 450 kOe are presented [15]. The hysteresis (blue shade in Fig. 2b), more clearly observed in T = 5 and 15 K, is attributed to a magnetic field induced pure structural transformation between the M and O(I) phases [15] (red and green regions in Fig. 2b).

In order to obtain the Curie temperature of the O(I) phase, $T_C^{O(I)}$, the same procedure developed in a previous work is employed [12]. The isothermal magnetizations are arranged in Arrott plots (see Fig. 3a) and in this representation, the expected behavior is linear according to Landau theory:

$$\frac{\mu_0 H}{M} = A(T - T_C) + BM^2 \qquad (6)$$

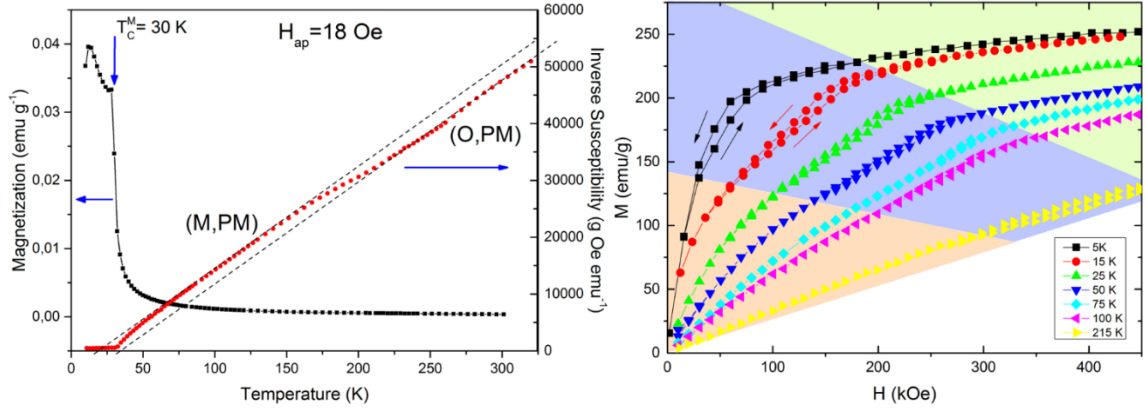

**Figure 2** (a) Magnetization and inverse of the magnetic susceptibility at 18 Oe [14]. (b) Magnetization isotherms of $Er_5Si_4$ measured in pulsed magnetic fields going up to 450 kOe [14].

where $A$ and $B$ are coefficients of the Landau free energy. Then, from the extrapolation of the magnetization of the low and high temperature phases, the inverse of the susceptibility and the spontaneous magnetization can be estimated, respectively (see inset of Fig. 3a). In this case, only the high temperature phase O(I) magnetization is of concern. Applying this method for every isotherm allows to obtain the spontaneous magnetization temperature dependence of the O(I) phase. These points were fitted to the Brillouin function to determine the saturation magnetization, $M_S$, and the molecular field parameter, $\lambda$. Finally, the O(I) magnetization as a function of temperature can be plotted and the corresponding Brillouin fitting estimates the Curie temperature value of $T_C^{O(I)} = 38.5$ K (see inset of Fig. 3b). It must be mentioned that the expansion of the Landau free energy here employed only goes up to the fourth degree, since it adequately describes the magnetic behavior of the system in the fitted data range.

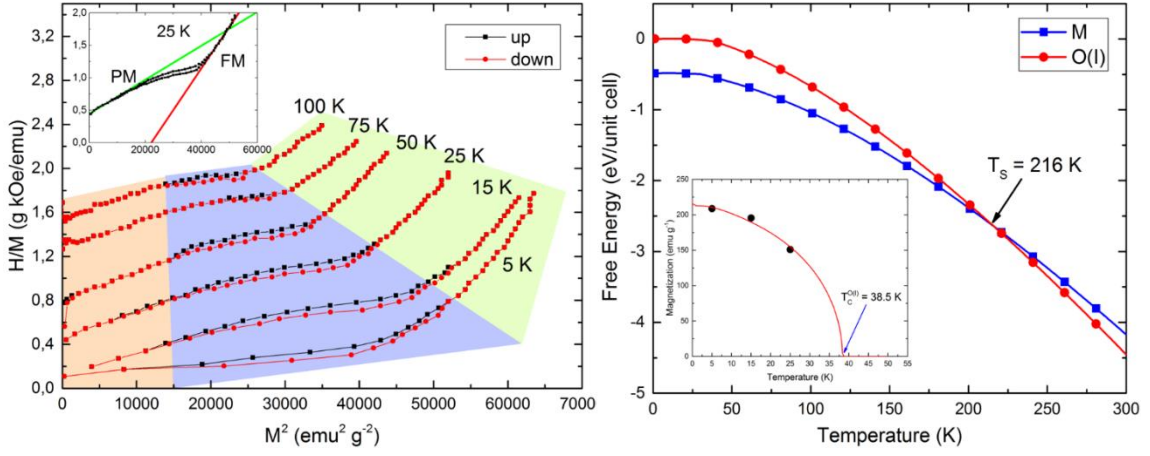

**Figure 3** (a) Representation of all the magnetization measurements in Arrot plots for increasing magnetic field (up) and decreasing magnetic field (down). (b) Temperature dependence of the free energy curves of the M and O(I) structures with no applied magnetic field. Inset: magnetization estimated from the extrapolation (black circles) and fitted Brillouin function (red curve) for the O(I) phase

It must be noted that there is some uncertainty to the estimate of $T_C^{O(I)}$ because the induced phase is not purely O(I), it is 61 vol.% O(I) and 39 vol.% M at H=50 kOe [15], and these factors affect the magnetic exchange interaction which in turn influences the Curie temperature estimation.

Despite the incompleteness transition, pressure studies [10] have found the magnetic transition temperature of the pure O(I) phase, $T_C^{O(I)}$, to be 37.5 K. Similar measurements were also performed for the three crystallographic directions in a single crystal, and both the magnetization and the $\Delta S^M$ temperature dependence reported a $T_C^{O(I)}$ of 35 and 37 K respectively [23], which corroborate well with the result obtained above, $T_C^{O(I)}$ = 38.5 K. Furthermore, from analysing the inverse susceptibility (see Fig. 2a) it is expected that $T_C^{O(I)} > T_C^M$ since $\theta_p^{O(I)} > \theta_p^M$.

## 5. Thermomagnetic model of phase transition

In order to estimate the expected temperature of the structural phase transition, the Debye temperature of the M structure, $\Theta_D^M$, was estimated to be 235.07 K using the approximation used in Ref. [27].

By computing the free energies of both phases according to Eq. (1), a structural transition is predicted at $T_S = 216$ K (see Fig. 3b) when considering a $\Theta_D^{O(I)} = 162$ K. It must be noted that the Debye temperature of the O(I) phase was chosen in order to obtain the experimental results. Despite not being reported, the expected values of the Debye temperatures are similar to the ones obtained for Tb$_5$Ge$_2$Si$_2$ and Gd$_5$Ge$_2$Si$_2$ [12, 28]. Furthermore, it was observed that in order to obtain a transition at 200 and 230 K, the required Debye temperature values were 156 and 167 K, respectively. This limits their range of values in order to be in agreement with experiment [4].

Despite the structural transition in Er$_5$Si$_4$ being in the range of 200-230 K, the $T_S$ obtained suggests that the model can indeed explain the transition in this compound as being a competition between two phases, the O(I) and the M. Using this model, the magnetic and lattice entropy variation (at $T_S$ with no applied magnetic field) were $\Delta S_M = 0$ eV/cell (since both structures are PM at ~200 K), $\Delta S_E \approx 0.079$ meV/cell , $\Delta S_L \approx 12$ meV/cell, thus, showing that the lattice entropy is the predominant element on the origin of the structural transition observed.

## 6. Conclusions

First-principles calculations verified the M structure as the stable phase at low temperature. This was only accomplished when including the Hubbard term (U=5.8 eV), typical of strongly correlated systems. Also, the Seebeck coefficient calculated from the DOS is in agreement with the value obtained experimentally.

The Curie temperature of the high temperature phase, O(I), was determined to be ~38.5 K which is in agreement with other values reported on the literature.

Finally, the thermomagnetic model was also used in the case of $Er_5Si_4$, which was shown to interpret the pure structural transition with a temperature of 216 K which is in good agreement with experiment. Nonetheless, advanced studies should be performed to have a more accurate Debye temperature estimation since the $T_S$ is greatly influenced by these parameters.

# Acknowledgements

This work was partially supported by the projects FEDER/POCTI n0155/94, PTDC/CTM/NAN/ 5414/2014 from Fundação para a Ciência e Tecnologia (FCT), Portugal. The authors thank the Portuguese Foundation for Science and Technology (FCT) for grants SFRH/BD/88440/ 2012 (JHB) and SFRH/BD/97591/2013 (MSB), SFRH/BD/ 97591/2013 (MBB), SFRH/BPD/82059/2011 (JNG), IF/01089/2015 (JSA). This work was developed within the scope of the project CICECO-Aveiro Institute of Materials, POCI-01-0145-FEDER-007679 (FCT Ref. UID /CTM /50011/ 2013), financed by national funds through the FCT/MEC and when appropriate co-financed by FEDER under the PT2020 Partnership Agreement. C.M. acknowledges the support of the Fundación ARAID. This work was supported by the Spanish Ministry of Science (through Project Nos. MAT2014- 51982- C2-R and C1-R, including FEDER funding) and the Aragon Regional government (Project No. E26).